# About the physical nature of some peculiarities of the primary cosmic radiation nuclei and gamma quanta spectra

T.T.Barnaveli*, N.A.Eristavi, I.V.Khaldeeva

Andronikashvili Institute of Physics, Tamarashvili 6, Tbilisi, Georgia
* E-mail: tengiz.barnaveli@gmail.com

## Abstract

About 20 years ago we published the data concerning some peculiarities of the behavior of cosmic radiation EAS **hadron component** spectra. The results pointed to the possible existence in the interstellar space of the background of weakly interacting objects of the mass (the energy of the resonance oscillations) of the order of 37 eV. On the other hand, the experimental data of the last years are pointing to the existence of cosmic **gamma** radiation with the specific spectrum having the steep right front again in the region of the order of 37 eV and the left front falling down to the energies of the order less than $10^{-6}$ eV. Obviously, no elementary object may possess such spectrum of frequencies or a decay spectrum. Such spectrum may have some certain system or construction consisting of many elements possessing their own resonance frequencies and together composing the spectra observed.

Further the possibility will be presented of exactly such explanation of the cosmic rays primary radiation spectra peculiarities experimentally observed. It is based on the hypothesis about the discreteness of the space and existence in it of the topological defects distributed with sufficient density. In the frames of the proposed model some essential experimental peculiarities of the primary cosmic radiation nuclei and gamma quanta spectra find the unified explanation.

## 1. The experimental data.

The data concerning some peculiarities of the behavior of primary cosmic radiation EAS **hadron component** spectra we have published in [1,2,3,4,5,6]. The archive material obtained by means of Tien-Shan high mountain installation was used for the analysis. The details can be found in the works quoted above. The results pointed to the possible existence of the background of weakly interacting objects of the mass (the energy of the resonance oscillations) of the order of 37 eV in the interstellar space. The interaction with this background could lead to disintegration of the primary cosmic radiation basic nuclei at some definite primary energies, specific for each kind of primary nucleus. The experimental eigenvalue of the background objects energy equals to 36.76±(10+5)%, where 10% is the error in estimation of the primary particle energy, and 5% is the statistical error. Further we will use the value 37 eV.

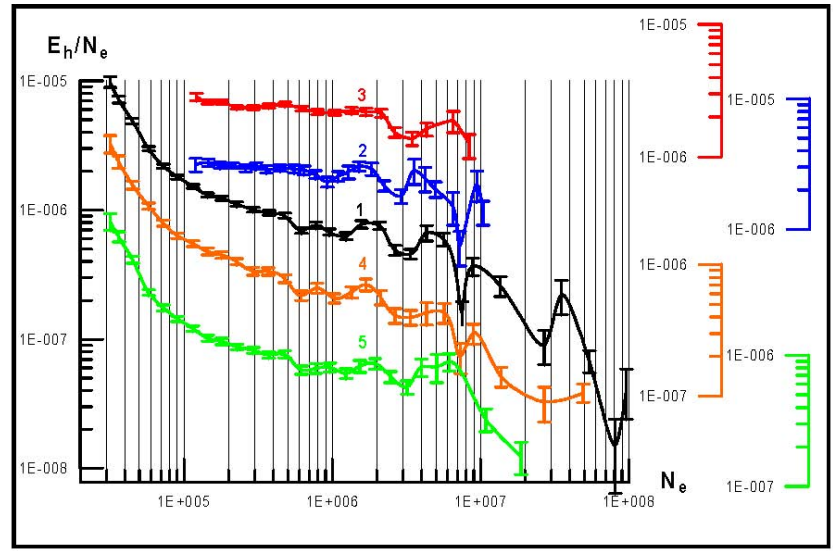

Fig. 1. The specific energies $E_h(N_e)/N_e$ of the EAS hadron component for the different configurations of apparatus and selection conditions. Details are given in the text and in [6]. The left vertical scale (arbitrary units) is given for the curve 1. The other curves are slightly shifted up or down to separate them from one another and to clarify the picture. The corresponding parts of the vertical scales are shown at the right edge of the figure.

In the Fig.1 one of the results [6] is shown, pointing to the existence of this phenomenon. The figure presents the specific energies $E_h(N_e)/N_e$ of EAS hadron component, registered in the multilayer hadron calorimeter of the area 36 m$^2$ being a part of the Tien-Shan complex installation. The data for the different configurations of installation and different selection criteria are shown.

Curve 1 – the hadron calorimeter, 16 layers, without target, the range of registration zenith angles $0^0 - 30^0$, the material was not subjected to preliminary selection – 37000 events;

Curves 4 and 5 – the same material divided into two parts according to the range of registration zenith angles $0^0 - 20^0$ (19000 events) and $20^0 - 30^0$ (18000 events)

Curves 2 and 3 – the calorimeter without target (8000 events) and with target (6000 events). The material was subjected to preliminary selection – the definite densities of particles were selected in separate groups of detectors.

The error bars allow for statistical errors. The errors of $N_e$ estimation are of the order of 10 %.

On the all curves in the region of $N_e > 2 \cdot 10^5$ one can distinctly see series of dips. The qualitative explanation of dip formation mechanism is given in [6]. In the same work the mutual attachment of the disintegrated nuclei kind to the dips on the spectra is given. Note that some kinds of the nuclei contribute to one and the same dip on the spectrum.

These irregularities most clearly are revealed in the region $N_e > 2 \cdot 10^6$, where one can easily trace the identical localization of these phenomena on all quoted curves, independently of the used experimental data, trigger conditions and selection criteria. Localization of these dips does not depend on event registration angle (curves 4 and 5). Here it is important to note that EAS were registered in the vicinity of the showers development maximum (0.7<S<1.3). So the change of the registration angle would not lead to the noticeable shift along the $N_e$ axis. As a matter of fact here we are dealing with the results obtained in several independent experiments.

.    It is interesting to consider the data [7], which are, in some sense, more detailed than that quoted in the figure above. The graph from this work is shown in Fig. 2.

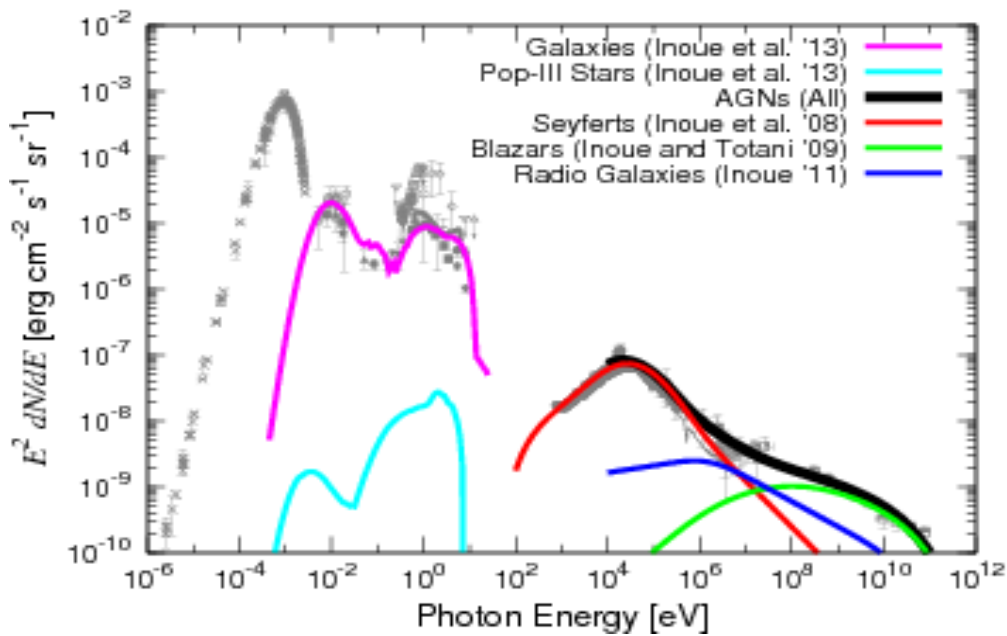

Fig. 2. The cosmic radiation from the different sources

To the problems considered in our article the left half of this graph has the immediate relation. The right fronts of the shown curves are located again in the energy region of the order of 37 eV. As in the Fig. 3, these curves are not continued in the region higher than 37 eV. We again pay attention to this important fact. Two of the curves have also strict left fronts and plateau with the local dips and ascends. These curves are so close to each other in many parameters (except for intensity), that the conclusion on their common **physical nature** is practically unambiguous.

The distinct indication of the special role of 37 eV energy one can see on Fig. 3 [8]. There is no smooth transition to the energies higher than 37 eV. The mechanism which contributes to the energies less than 37 eV does not exist higher than 37 eV. In the vicinity of 37 eV the spectrum shows the kink, here is the singular point. Note that on this graph the whole hump to the left of 37 eV lasts down to ~$10^{-6}$eV.

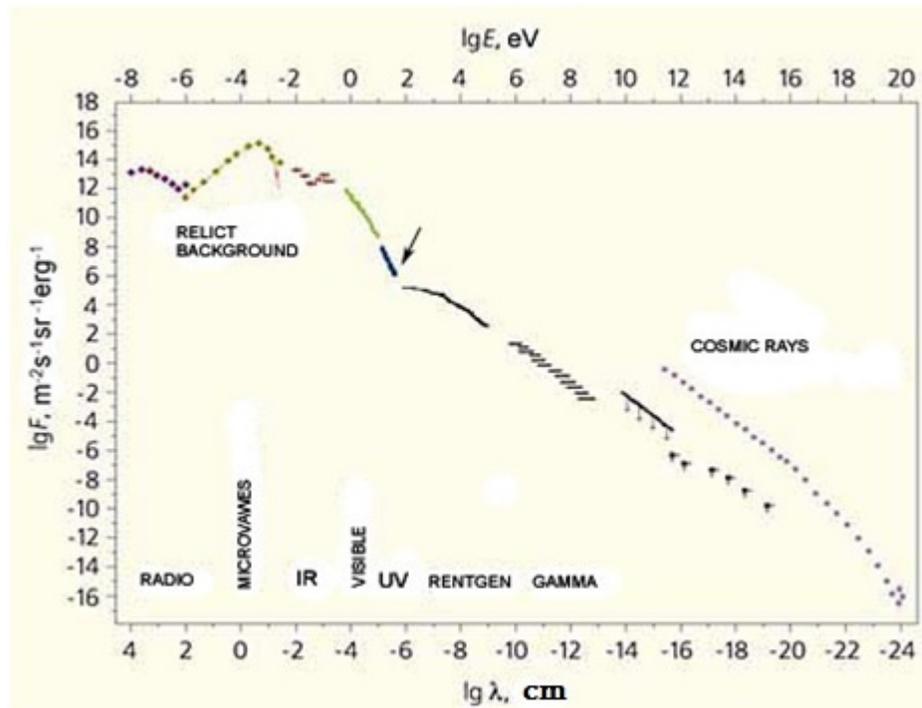

Fig. 3. The primary gamma quanta unified spectrum.
The arrow corresponds to the energy 37 eV.

The absence of such background particle creation in accelerator experiments may mean, that it is an object of a different nature – for example the topological defect of the space.

## 2  Discussion.

### 2.1. The work hypothesis

Below the possibility of such explanation of the observed peculiarities, based on the hypothesis about the discreteness of the space and existence in it of topological defects distributed with sufficient density is discussed.

The possible discreteness of the space structure and the essential role of the defects of this structure are discussed for enough long time. Among the most important recent works in this field let us mention the works of S.Hossenfelder, in particular [9,10,11], where many important results in this field are cited as well.

Our main work hypothesis here is as follows. The space consists of the cubic cells with the Plank size ribs, i.e. of the order of $1.6 \cdot 10^{-33}$ cm – it is the basic state of the space cell. The cells compose the structure related to the cubic symmetry. Each cell has the 6 closest neighbors. The cubic shape of the cells provides their really dense packing and, at the same time, the high degree of isotropy and locally Euclidean geometry. Of course the cells allow some deformations (bends, twists, compressions, etc.) providing the possibility of space curving. Due to extremely small dimensions the space cells are the quantum objects and correspondingly undergo the quantum tremble and fluctuations – further the corresponding considerations will be quoted.

The contradictions emerge at consideration of space generation and expansion process, at all stages of the universe development – from the "big bang" and including the present time. The point is that the space creation and expansion must obviously obey to the spherical symmetry – in any case at very early stages, **until the role of gravitation from the created matter is too small to cause the essential curvature of space.** This means that the construction **of cubical inner symmetry** must create the system which **outwardly is developing as a spherical system.** At expanding of the sphere each part of its surface is expanding in all directions along the surface. This expansion must lead to creation of the new spatial cells. At the same time the pressure existing inside the system (see further about this) tries to retain their densest packing. As a result, the division of each arbitrary large circle of the developing sphere (so the picture of what is happening gets more obvious) into the sectors necessarily takes place. Really the division of the space into pyramids with their tops located in the centre of the system takes place. Besides, the sector of the densest package of the spatial cells is separated from the nearby laying analogous sector by the "vacant" gap, **which compensates the difference of the orientation angles of these sectors.** Exactly these gaps in the environment of spatial cells are forming the topological defects which will be considered further. In the scheme in Fig. 4 the part of large circle of such expanding sphere lying in the plane of the arbitrary sectors radial edge is shown.

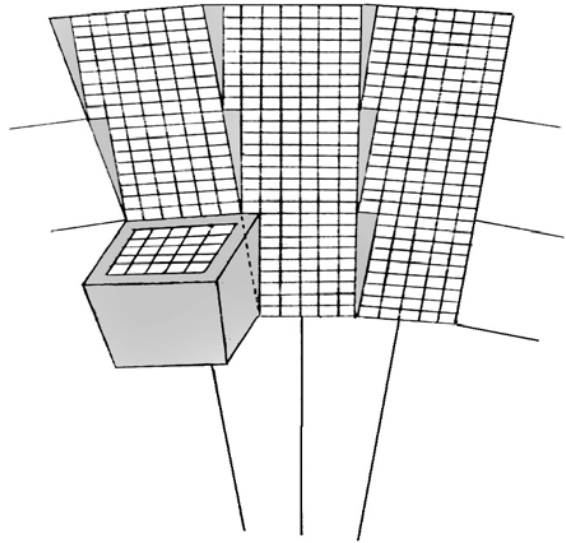

Fig.4 .The schematic sketch of space structure and topological defects - the separate area of a large circle. The spatial cells are shown as a white squares.

The numerical evaluations of the shapes and dimensions of these defects will be given further. In our model the angles between the sectors, and evidently the width of quantum state of the construction parts depend on the dimensions of the cells. **The shape of the cells inside the vacant gaps is already not the cubical.** This may require the assumption of the possible existence of a rather broad range of cell deformations. **This (and it is very important) leads to the anisotropy of space in the region of defects which defines the possibility of cosmic radiation scattering on the defects.**

As we have noticed from the very beginning, the considered construction must have the spherical symmetry, in any case at the first stages of development. Of course the forces of gravitation later inevitably lead to the curving of the space. But **locally** the space remains to be flat and retains its inner structure, as in fig. 4.

At all stages of expansion this structure provides the possibility to "build in" (to insert) a new cell into the vacant gap. In reality, due to quantum "tremble" of the dimensions and shape of the inserted cells (until they become the members of sector structure) they can be born in these vacant gaps in different quantities in order to be then squeezed off by the new born cells and built in the structure of sector.

The "vacant gaps"– the topological defects in the schema in Fig. 4 are darkened and have a triangular shape in the transverse cross section. In reality they are prisms with triangular bases directed along the radiuses. Using the terminology of the work [9] these defects may be regarded as so called "local" defects. Their properties, in particular, the influence on the isotropy of space and on Lorenz invariance of physical processes are analyzed in details in [9] and in the works cited there. We will touch these questions below.

The transverse dimension of the separate "standard" space cell is larger than the maximal transverse section of the vacant gap. The last words in respect to the vacant gap must be understood in the sense that one can`t place there the standard cell until the gap becomes broad enough in the process of general expansion. After this, the gap becomes blocked by the standard cell, behind which a new radial row of the cells starts to grow. So, at each new stage of development the totality of the space cells of each sector becomes broader by one radial row.

It is obvious that the construction shown in Fig.4 must have approximately the equal size both in wide and in depth. As an example this is shown in the left bottom sector in Fig. 4. So, the directions in "wide" and in "depth" can`t be mutually normal. However, the number of sectors in any cross section is so large ($\sim 10^{27}$ - the evaluation will be given below) that the difference from the mutual normality will be of the order of $10^{-27}$, so by consideration of the construction by way of several neighboring sectors one can neglect this value.

So, the defects as a whole form the system of "boxes", but they have neither the "lids" nor the "bottoms". The **radial** dimensions and the shape of vacant gaps everywhere are approximately the same while their orientation is always along the radius of the corresponding large circle. So the defect radial dimensions are of the same order at all stages of expansion. Exactly this dimension defines the characteristic resonance frequency (dominant mode) of defect oscillations – about 37 eV.

The defect dimensions **normal to radii** increase with the stages of expansion. These dimensions, of course, define also the corresponding modes of oscillations – practically continuous spectrum of low energies down to the energies of relic background and less, of the order of $10^{-6}$- $10^{-7}$eV (evaluations see below).

The spatial cells are accumulated with one edge oriented towards the system center. Here it is very important to underline, that the oscillations of defects are excited due to the interaction with the environment particles – the different kinds of cosmic radiation and heat radiation. **So, these oscillations of defects do not contribute additionally to the average density of universe energy.** (Unless with the exception of the contribution of some local tension of space construction near the defects). These oscillations are again irradiated back in space as gamma quanta of the corresponding energy.

These topological defects really play their very important role in the process of space expansion – they are the natural regions of the new cells generation **inside the already formed volume of space**, as soon as in the process of system expansion such possibility and necessity emerge.

## 2.2. The mechanism of construction formation

After the hypothetical scheme given above let us try to outline the possible mechanism of the formation of such structure of space. The main question is – may the phenomenon like Bose condensation be a part of big bang and universe expansion? Really, the Plank space cells doubtless are the quantum objects. In this case the spin 0 obviously has to be ascribed to them; otherwise it may lead to very undesirable properties of the space. Thus it follows that they must obey to Bose statistics. But **for this, the existence of the physical space with its properties and symmetry is necessary**. In the giant fields of big bang with the necessity the birth of the particles of any kind out of the vacuum is taking place, including the space cells. Behind the temperature decrease at some moment the conditions emerge for creation of the areas (domains) of full-bodied space, possessing all physical laws including the symmetry. In other words the phase transition takes place, analogous to crystallization process. Here it gets possible the birth of the integer spin particles out of the vacuum in the form of Bose condensate [12, 13, 14, 15]. In particular the further birth of the spatial cells may take place in the similar way. In the course of this process there arises the possibility to minimize the inner energy of the main construction caused by the chaotic location of the gaps (i.e. the defects) between the space cells. The process of gradual minimization of the inner energy leads to the construction shown in Fig. 4. Really, the substance born in the process of big bang creates the gravitational field of the central symmetry. The optimal way to minimize the inner energy of space construction is to divide the space into sectors (see schema in the Fig. 4). Each sector will have its own orientation of the space cells.

**Note that the cells in each sector, i.e. of the same orientation regarding the system center, are in the same quantum state.** All the other quantum numbers (if the cells possess them) are taken to be the same. In each separate sector the Bose statistics provides the densest package of the cells. **Just the Bose statistics creates in the system the negative pressure and thus a returning force in case of the mutual shift of the cells. In other words in this model the physical space is a Bose condensate of Plank cells.** Just this fact provides the behavior of each wall of defect (of "box") **as a single whole** at the interaction with the different kinds of radiations of environment.

The cells of the given orientation with regard to the center of the system can be created simultaneously or in different times in many places of the developing universe. However, the regular sequence of cells orientation, like in Fig. 4, is evidently the most advantageous one from the energetic point of view. In such construction the topological defects obviously will be less in total volume and will have more regular, smooth shape. I.e. with increasing number of development stages (of time) only the situation close to that shown in Fig. 4 will be fulfilled and survived. It will be predominant starting from some period of Universe development.

The previous, old stages may have a chaotic structure. Besides, the process of Bose condensation inevitably acquires the multichannel character providing each sector with the cells of needed orientation. This may take place at very high fields and scales of big bang. The processes of big bang provide the cells of all these orientations and **in amounts able to stimulate the further development of Bose condensates in all sectors and at all stages of structure development**. Just this process we have denoted above as the multichannel Bose condensation.

The radial symmetry is defined by the gravitational or some other calibration field, the argument of which is the orientation of the cells. As a result the space gets divided into sectors each of which has its own cell orientation.

The number of sectors evidently is defined by scattering of cell orientation angles, i.e. by the "width of cell state "$\Delta\theta$, i.e. by uncertainty relation $\Delta\theta \cdot \Delta M \sim \hbar$. The width of state changes depending on the temperature inside the system. With decrease of the temperature the spread of orientation angles of the parts of construction decreases, until it reaches the level defined now mainly by the construction of the defect itself. The further decrease of the temperature does not change anymore the width of state. The situation is the same in case of the sectors. At the early stages the sectors do not exist at all or exist as the domains scattered chaotically. Later the sectors emerge and increase in number in accordance with the decrease of temperature inside the system, until the width of state reaches the level defined by the inner quantum state of the construction. At this stage the width of state, i.e. the level of cell orientation spread gets smaller than the angle between the sectors. This takes place at temperatures of the order of $6 \cdot 10^6$ degrees, corresponding to the energy of 37 eV. Evidently from this temperature the "quiet" stage of universe development is starting. This temperature is exactly that one, which provides the stability of described construction, with the given relative dimensions of each separate cell and geometry of the whole construction.

Let us quote some numerical evaluations.

As it was noticed above the defect radial sides (see fig. 4) **have practically the same length at all stages of expansion.** Its spread obviously depends on the quantum tremble of the space cells inside the defects. If one assumes that the characteristic frequency of defect oscillations (the main mode) 37 eV corresponds to this dimension, it is easy to see (starting from $\Delta p \cdot \Delta x \sim \hbar$) that it must be of the order of $5.4 \cdot 10^{-7}$ cm. From Fig. 4 it is clear that each wall of defects under consideration has the shape of a triangle prism. The maximum thickness of this wall is of the order of $1.6 \cdot 10^{-33}$ cm (Plank length). The rows parallel to the radii contain about $5.4 \cdot 10^{-7} / 1.6 \cdot 10^{-33} = 3.375 \cdot 10^{26}$ Plank lengths. These figures define as well the value of the angle between the sectors of the large circle: $1/3.375 \cdot 10^{26} \sim 10^{-27}$ rad. The radius of the universe R = $1.3 \cdot 10^{28}$ cm (for the universe age 14 · *10⁹* years). The number of rows (micro stages of development 5.4·*10⁻⁷* cm each*)* will be 1.3·*10²⁸*/ $5.4 \cdot 10^{-7} = 2.4 \cdot 10^{34}$. Each stage adds 1 cell perpendicular to the radius in each sector. **So, exactly this number of Plank lengths at the recent stage contains the side of the defect normal to radius.** Correspondingly, the recent lengths of the sides normal to the radii are $2.4 \cdot 10^{34} \cdot 1.6 \cdot 10^{-33} \approx 40$ cm. Naturally we`ll obtain the same, starting from R and the angle at the vertex of defect, i.e. the angle between the sectors: $3 \cdot 10^{-27} \cdot 1.3 \cdot 10^{28} \approx 40$ sm. The corresponding resonance energy is $4.99 \cdot 10^{-7}$ eV.

Of course one can object here that formation of the structure by stages was started not from the moment of big bang, but much later, when the temperature of the environment decreased enough. But even if one admits that it started after1 $5 \cdot 10^6$ years from big bang, the correction will be not more than 0.001, that is not discoverable for today.

So 40 cm is in fact the order of defect dimension **along the normal to radius for the recent stage of expansion**. One can say conventionally that the **flat** square of the side 40 cm on the recent surface of the sphere with its center at the joint of four adjacent "boxes" contains normally to radius one defect in average. Each centimeter of the length of each of 4 walls of defect "box" is equivalent by volume of $0.5 \cdot 5.12 \cdot 10^{-66} \cdot 5.4 \cdot 10^{-7}$ cm³ (0.5 - reflects the median thickness of the defect wall, 5.12 is 1.6³, 5.4·10⁻⁷ is the "height" of the «box»). So taking into account that the length of the rwalls of the defect inside the square under consideration 40 · 40 cm² and the depth of consideration 1 cm, we obtain for the recent stage of Universe development that each defect occupies the volume $0.625 \cdot 10^{33} \cdot 40$ Plank units ( $0.625 \cdot 10^{33}$ is the number of Plank lengths in 1 cm). Dividing by the volume of parallelepiped 40· 40 · 1 cm³ we obtain $0.625 \cdot 10^{-33} \cdot 40 / 1600 \cdot 10^{99} \approx 00156 \cdot 10^{-67}$, i.e. of the order of $10^{-69}$ 1 per cm³, or of the order of $1 \cdot 10^{-17}$ per fm³. It is much less than the experimentally acceptable upper limit <1 fm⁻⁴ of spatial volume to keep the Lorentz invariance of the physical lows quoted in [9]. Here one must as well take into account that the "defect" cells are not spread around the volume, but compose the compact structure of the defect wall.

The number of defects per 1 section of each regular spherical layer by the large circle plane (i.e. within the limits of each micro stage of expansion by $5.4 \cdot 10^{-7}$ cm) must be one and the same everywhere to compensate the same angle 2π, differing the cubical symmetry from spherical one. Note that the number of defects per one unit of space may serve as a **measure of the development phase of the given region of universe.**

Let us make some additional notes. 37 eV is the basic, most energetic self- oscillation mode of the defect. These oscillations can be excited only by interaction with some outer cosmic radiation as we have mentioned above. This mode is fully defined by the minimal dimensions (the radial length) of defects construction. However, in fact the defects are all the walls of the "boxeous" structure. The walls which are normal to radii have essentially larger dimensions, especially at the "recent" stage. These dimensions compose the practically continuous spectrum resulting in the correspondingly continuous spectrum of energies less than 37 eV down to ≤ $10^{-7}$ eV. Actually each of these boxes is a resonator. **So the space as a whole is a giant system of resonators.**

If the plank length was not changed in course of Universe development then all basic resonance frequencies remain unchanged. The continuous spectrum of dimensions provides the resonance at broad range of frequency. That, if not being irradiated back- gets gradually distributed among the whole system of resonators. However, if the **re- radiation** of 37 eV gamma quanta (the basic frequency) took place, they of course were subjected to cooling in the process of Universe expansion, like the relic radiation. So most likely there should exist the spectra of photons occupying the gap $10^{-7}$ – 37 eV, i.e. exactly the range provided by the whole entity of the topological defects. As we noticed above, there is no contribution of this process to the right side of 37 eV, here the contribution stops. The distinct notation of this one can see on Fig. 4 [8]. The mechanism which contributed to the energies less than 37 eV does not exist higher than 37 eV. In the vicinity of 37 eV the spectrum shows the kink, here is the singular point. Note that on this graph the whole hump to the left of 37 eV lasts down to ~$10^{-7}$ eV, i.e. it really occupies just that range which is provided by the whole totality of the topological defects according to the evaluations made above. Here, the overlap of the radiation from defects and the relic radiation takes place.

Let us return to the data [7]. The graph from this work was shown in Fig. 2. As we noticed above the left half of this graph has the direct relation with the questions considered in our article. The right fronts of the curves shown are located again in the energy region of the order of 37 eV. These curves again are not continued in the region higher than 37 eV. Everything is going to show that we are dealing again with the radiation from the same topological defects.

Let`s discuss it in more details. To the right fronts and right parts of the spectra contribute mainly the elements of the construction located **along** the radii, while to the left fronts and into the left abatement down to $10^{-7}$eV contribute the structures located **normally** to radii.

The construction completely corresponds to the characteristic maximum frequency (i.e. to the energy 37 eV) and provides the stability at the temperatures~ $6·10^6$K. The left front of the third curve goes down to the energies of the order of $10^{-6}$eV, that corresponds to the quantitative evaluations quoted above. Thus this curve represents the radiation of the basic structure of the topological defects.

Some difficulties are caused by existence of plateau on the spectra, more exactly, not the plateau itself, but the existence of the second and even the third peak in the very left part of the plateau. Most probably these are the resonance frequencies not of only one wall but of collective oscillations jointly of two and even of three walls of defect (for the very left curve). It is clear that the radial, i.e. most energetic modes of oscillations are meant.  The numerical evaluations of the defect oscillation spectra must be done on the basis of construction (plates, boxes) oscillation theory. Though probably there is a lack in input data. May be it will be worthy to carry out the computer model experiment with the different input data, i.e. to solve the reversed task.

Here however one has to note another very important moment. Of course **inside the very hot objects** the described construction is subjected to very intensive irradiation. However, the gamma quanta scattered or re-radiated from the structure inside of such objects are undergoing the many times scatter or re- radiation from the matter inside the hot object and lose practically whole their energy.  So, the "original" quanta are radiated into outer space really only from the border regions of such objects. So, their intensity is low and the region of the very small energies on these curves may not be seen due to very small experimental statistics.

## 3. Conclusions

In principle such topological defects could be discovered by means of the special accelerator experiment at very low energies and maximally precise measurements. In such low energetic reactions the part of the energy may be spent on the excitation of these objects and the "missing* energy may emerge. The presence of "missing" energy (if it will be) will point to the existence of resonance phenomena at the corresponding frequencies.

In the course of such experiment it could be possible to evaluate the maximum size (i.e. the minimal frequency of self oscillations) of defect parts located normally to the radii of large circles at the present stage of expansion. This **will make it possible to evaluate the number of sectors in the large circle and hence to obtain the experimental evaluation of the Planks length, i.e. of one of the fundamental world constants.** The difficulty of such experiment is the necessity to provide the required precision of the measurements with the account of kinematics of the observed interactions.

To conclude, let us note that in the frames of the model described the essential peculiarities of the primary cosmic radiation of nuclei and gamma quanta spectra find the unified explanation.

**Acknowledgements**           The authors express their sincere gratitude to  O.V.Kancheli , to G.I.Japaridze  and to A. A. Khelashvili  for good-wishing critical remarks and valuable discussions. We are sincerely thankful to I.R.Lomidze for discussion and valuable advises.

References

1. T.T.Barnaveli, N.A.Eristavi, I.V.Khaldeeva, Z.Shergelashvili. Phis. Lett., B 346 (1995) 178.
2. T.T.Barnaveli, T.T.Barnaveli (jr), A.P.Chubenko, N.A.Eristavi, I.V.Khaldeeva, N.M.Nesterova, Yu.G.Verbetsky. Phis. Lett., B 369 (1996) 372.
3. T.T.Barnaveli, T.T.Barnaveli (jr), N.A.Eristavi, I.V.Khaldeeva, Yu.G.Verbetsky. Phis. Lett., B 381 (1996) 307.
4. T.T.Barnaveli, T.T.Barnaveli (jr), N.A.Eristavi, I.V.Khaldeeva, Yu.G.Verbetsky. In Very high Energy Phenomena in the Universe . Rencontres de Moriond. (1997) 419.
5. T.T.Barnaveli, T.T.Barnaveli (jr), A.P.Chubenko, N.A.Eristavi, I.V.Khaldeeva, N.M.Nesterova, Yu.G.Verbetsky. Arxiv. Astro-ph/ 0208275 (2002).
6. T.T.Barnaveli, T.T.Barnaveli (jr), N.A.Eristavi, I.V.Khaldeeva. Arxiv. Astro-ph. 0310524. V 1. (2003).
7. Yoshiuki Inoue. ArXiv: 1412.3886 v1 [astro-ph. HE], (2014).
8. B.A.Khrenov, M.I.Panasiuk. Priroda, 2 (2006). In russian,
9. S.Hossenfelder. ArXiv: 1401.0276 v1 [hep-ph], (2014).
10. S.Hossenfelder. ArXiv: 1309.0311 v2 [hep-ph], (2014).
11. S.Hossenfelder. ArXiv: 1309.0314 v2 [hep-ph], (2014).
12. A.B.Migdal. JETP **61**, 2209 (1971).
13. A.B.Migdal. UPN **103** , 369 (1977).
14. Ya.B.Zeldovich, V.S.Popov. UPN **105**, 6, (1971).
15. T.T.Barnaveli, N.A.Eristavi, I.V.Khaldeeva, Arxiv: astro-ph . HE/ 1604.04152 (2016).